# Quantum Hall Effect at 0.002 T

**Alexander S. Mayorov[1]†, Ping Wang[1,2]†, Xiaokai Yue[3]†, Biao Wu[1], Jianhong He[4], Di Zhang[1], Fuzhuo Lian[1], Siqi Jiang[1], Jiabei Huang[1], Zihao Wang[5], Qian Guo[6,7], Kenji Watanabe[8], Takashi Taniguchi[9], Renjun Du[1], Rui Wang[1], Baigeng Wang[1,10,11]\*, Lei Wang[1,10,11]\*, Kostya S. Novoselov[5]\*, and Geliang Yu[1,10,11]\***

**Affiliations:**

[1]National Laboratory of Solid State Microstructures, School of Physics, Nanjing University, Nanjing 210093, China

[2]College of Physics and Mechatronic Engineering, Guizhou Minzu University, Guiyang 550025, China

[3]Department of Physics, Sichuan Normal University, Chengdu, Sichuan 610066, China

[4]College of Physics, Sichuan University, Chengdu, Sichuan 610064, China

[5]Institute for Functional Intelligent Materials, National University of Singapore, Singapore

[6]Department of Physics and Astronomy, University of Manchester, Manchester, UK

[7]National Graphene Institute, University of Manchester, Manchester, UK

[8]Research Center for Functional Materials, National Institute for Material Science, 1-1 Namiki, Tsukuba 305-0044, Japan

[9]International Center for Materials Nanoarchitectonics, National Institute for Material Science, 1-1 Namiki, Tsukuba 305-0044, Japan

[10]Collaborative Innovation Center of Advanced Microstructures, Nanjing University, Nanjing 210093, China

[11]Jiangsu Physical Science Research Center, Nanjing 210093, China

†These authors contributed equally to this work.

\*Authors to whom any correspondence should be addressed: bgwang@nju.edu.cn, leiwang@nju.edu.cn, kostya@nus.edu.sg and yugeliang@nju.edu.cn

# Abstract

Graphene enables precise carrier-density control via gating, making it an ideal platform for studying electronic interactions. However, sample inhomogeneities often limit access to the low-density regimes where these interactions dominate. Enhancing carrier mobility is therefore crucial for exploring fundamental properties and developing device applications. Here, we demonstrate a significant reduction in external inhomogeneity using a double-layer graphene architecture separated by an ultra-thin hexagonal boron nitride layer. Mutual screening between the layers reduces scattering from random Coulomb potentials, resulting in a quantum mobility exceeding $10^7 \, \text{cm}^2 \text{V}^{-1} \text{s}^{-1}$. Shubnikov–de Haas oscillations emerge at magnetic fields below 1 mT, while integer quantum Hall features are observed at 0.002 T. Furthermore, we identify a fractional quantum Hall plateau at a filling factor of $\nu_{\text{tot}} = -10/3$ at 2 T. These results demonstrate the platform's suitability for investigating strongly correlated electronic phases in graphene-based heterostructures.

# Introduction

Graphene, a single layer of carbon atoms arranged in a hexagonal lattice, has emerged as a cornerstone for exploring two-dimensional (2D) electronic systems.(*1-8*) Its discovery and subsequent investigations have unveiled many unusual electronic properties,(*3*) such as high carrier mobility, ballistic transport at micrometer scales,(*9*) and the quantum Hall effect at room temperature.(*10*) These attributes stem from graphene's unique band structure, characterized by Dirac cones where the conduction and valence bands meet at the Dirac points, giving rise to massless charge carriers with linear dispersion.(*11*, *12*) However, realizing the full potential of graphene's electronic properties in practical devices has been hindered by challenges related to environmental sensitivity, substrate-induced disorder, and charge inhomogeneity, which mask its intrinsic behavior.(*13*, *14*) Our research offers a transformative solution to these challenges, paving the way for practical applications in electronic devices.

The dominant scattering mechanism in graphene is long-range Coulomb scattering induced by charge impurities, which form electron-hole puddles that indicate charge inhomogeneity near the charge neutrality point (CNP).(*3*, *14*, *15*) Encapsulation of graphene with hexagonal boron nitride (hBN) and the use of dual graphite gates have markedly improved the quality of graphene-based samples.(*16*) As internal sample disorder is reduced, the impact of boundary disorder becomes increasingly evident. In Hall bar devices, mobility is directly proportional to channel width.(*7*, *17*, *18*) Edgeless samples exhibit enhanced performance, such as Corbino devices(*19*, *20*) and Hall bars patterned by electrostatic gating.(*21*) Substrates with a higher dielectric constant can more effectively screen long-range Coulomb scattering, enhancing mobility.(*22*)

Recently, a remarkable result has been reported in monolayer graphene encapsulated in hBN with a graphite proximity gate separated by an ultrathin hBN dielectric. A nearby graphite gate screens long-range potential fluctuations, allowing for transport mobility of the order of $10^8 \, cm^2 V^{-1} s^{-1}$ and residual charge inhomogeneity of $3 \times 10^7 \, cm^{-2}$.(39) The onset of Shubnikov-de Haas (SdH) oscillations has been observed at low magnetic fields of 1–2 mT. Nevertheless, the presence of a proximity gate reduces the gaps for fractional quantum Hall states by a factor of 3–5 times.(39) Large-angle twisted bilayer or trilayer graphene encapsulated in hBN has revealed tunable Coulomb screening and the ability to achieve quantum

mobility of approximately $2 \times 10^6$ cm$^2$V$^{-1}$s$^{-1}$.(40)

Beyond the previously mentioned factors for enhancing mobility, a notably more effective approach is to employ a decoupled double-layer graphene configuration. As the literature demonstrates,(2) hBN flakes with thicknesses ranging from 4 to 12 nm are used to separate two graphene layers, each equipped with independent probing electrodes. Increasing the carrier concentration in one graphene layer enhances the screening of charge impurity scattering at zero magnetic field, thereby increasing the resistivity in the other layer, which can exceed $\rho_q = h/4e^2$. This threshold indicates localization near the CNP for a monolayer, primarily caused by the dominance of short-range intervalley scattering within this range.(23–25)

When the hBN spacer thickness is reduced to 2–5 layers, the ohmic contact electrodes connect to the two graphene layers. We observed that even though both graphene layers are near the CNP, the mutual screening effect markedly enhances the charge carrier uniformity of these devices. Mutual screening refers to the reduction of scattering in one graphene layer due to the presence of a second graphene layer, which effectively shields the electric fields from impurities. As a result, impurities primarily affect only one graphene layer, and their electric fields do not penetrate sufficiently to scatter charge carriers in the opposite layer. This effect has not been extensively studied before.

Our structure, consisting of graphene/few-layer hBN/graphene, is a double-layer graphene system (DLG). Hall edge states in two-dimensional samples are typically observed under strong or quantizing perpendicular magnetic fields. Here, we report measurements of the quantum Hall effect (QHE) performed under a magnetic field of 0.002 T and the fractional quantum Hall effect (FQHE) at 2 T in the DLG.

## Results and Discussion

### Basic characteristics

We have fabricated several devices, D1, D2, D3, D4, D5, and DG2, with different widths and types of contacts. As shown in Fig. 1(a), a few hBN layers are intercalated between two graphene layers, encapsulated by thick films of hBN and graphite. There are two types of ohmic contacts: Type I and Type II. Type I is the conventional 1D edge contact, featuring direct contact between the metal and the graphene layers.(7) Type II utilizes a thick graphite layer to connect the metal electrodes and graphene, thus avoiding the disorder introduced during metal deposition and local charge doping due to work function disparities.(*26*) Furthermore, Type II avoids the strong p-n junction formation in Type I, stemming from misalignment between the top gate and the channel. The inset of Fig. 1(b) shows the optical image of device D2, where the ohmic contact is Type II. Furthermore, D2 ($5.5\,\mu m$) has a wider channel width than D1 ($1.5-2\,\mu m$). To identify the most critical factors affecting device quality, we fabricated two additional devices, D3 and D4 (Supplementary Fig. 1). D3 and D4 utilize the Type I ohmic contact, and D4 ($4\,\mu m$) also features a wider channel width. Our best-performing device, D5, utilizes graphite contacts.

Figure 1(b) shows the $R_{tot}(V_{tg}, V_{bg})$ for the hBN (2L) region of device D1 (see also Supplementary Fig. 2 for comparison). The resistivity reaches its maximum near $D = 0\,V/nm$ in contrast to drag samples with independently connected layers, due to strong e-e interactions (Supplementary Fig. 3 and Supplementary Note 1 present DG2 sample data). The resistance peak splits, indicating the CNP of the upper (dashed green line) and lower (dashed red line) graphene layers.(*4, 27, 28*)

In our DLG structure, the interlayer charge tunneling is effectively suppressed by the hBN spacer, allowing the two graphene layers to be treated as parallel transport channels. Each layer maintains the intrinsic properties of monolayer graphene. Consequently, a pronounced maximum in resistivity near $V_{tg} = 0\,V$ and $V_{bg} = 0\,V$ is attributable to the diminished carrier density in both layers. As $D$ increases, it injects electrons into one layer and holes into the other, or vice versa. Because of near electron-hole symmetry in graphene, this results in a decrease in total resistivity as both layers become conductive.(*5, 27–30*)

Graphene has a remarkable linear dispersion relation near the Dirac points, which causes the absence of backscattering, making it very different from conventional high-mobility 2D electron/hole gases based on GaAs/AlGaAs heterostructures. In a remarkable paper (41), the electronic properties of graphene naturally decoupled from bulk graphite were investigated at small magnetic fields. The formation of Landau levels revealed very high electronic quality in these naturally occurring graphene layers, where the quantum mobility exceeded $2\times 10^7\,cm^2V^{-1}s^{-1}$. This result was attributed to the ultralow carrier density and the absence of extrinsic scattering sources. Therefore, the main task in achieving ultra-high mobility in graphene is to restore its intrinsic high quality by improving fabrication methods. To provide insight into the mechanism by which our double-layer approach enhances mobility, we consider two graphene layers separated by a thin hBN layer of thickness $d$. The bare intralayer and interlayer Coulomb interactions are given by $V_{11} = V_{22} = 2\pi e^2/\kappa q$ and $V_{12} = V_{21} = (2\pi e^2/\kappa q)e^{-qd}$, respectively, where 1 and 2 denote the upper and lower layers, $q$ is the transfer momentum, and $\kappa$ is the dielectric constant of the hBN layer. The interlayer scattering of electrons is assumed to be negligible due to the insulating nature of the hBN separation. Nevertheless, the two graphene layers screen each other, effectively reducing the intralayer Coulomb scattering at zero magnetic field (see Supplementary Fig. 4 and Supplementary Note 2). The scattering time $\tau$ determines the mobility, given by $\mu = \sigma/ne = 2ev_F\tau/hn$, where is the conductivity, $v_F$ is the Fermi velocity, and $n$ is the concentration. Due to the enhanced screening, both the scattering time $\tau$ and the mobility $\mu$ increase. In Fig. 1c, we calculate and plot the ratio of mobilities for the double- and single-layer structures. The mobility enhancement becomes more pronounced for smaller layer separations $d$, with the ratio reaching values as high as 3 to 4, which aligns well with the high mobility observed in our double-layer samples.

**Integer Quantum Hall Effect**

We now proceed to the main result of our paper. The device quality can be significantly improved by increasing the channel width to more than 6 mm. For a wide range of gate voltages not very close to the Dirac Point (DP) $n_{tot}$ and $D$ can be defined as: $n_{tot} = (C_{bg}V_{bg} + C_{tg}V_{tg})/e - n_0$ and $D = (C_{bg}V_{bg} - C_{tg}V_{tg})/2\varepsilon_0 - D_0$, where $C_{tg}$ and $C_{bg}$ are the top and bottom gate capacitances per unit area, respectively. $n_0$ and $D_0$ are the residual doping and residual displacement field, respectively. $e$ is the elementary charge, $\varepsilon_0$ is the vacuum permittivity. Fig. 2(a) shows the transverse resistance of device D5 in the $n_{tot} - D$ space at $B = 5$ mT and $T = 20$ mK. The plateau-like structure clearly indicates the observation of the QHE at a low magnetic field of 5 mT. Figures 2(b) and 2(c) show longitudinal and Hall resistance as a function of magnetic field and total concentration. Two cuts are shown in Figs. 2d and 2e. The plateaus at total filling factors of -4 and +4 begin to appear at a magnetic field of 2 mT (Fig. 3(d)). This point can be taken as the onset of the QHE for our device D5, coinciding with a zero-resistance state in the corresponding longitudinal resistance. The sequence of quantum plateaus observed is consistent at $B = 5$ mT with the $4(N + 1/2) \times 2$ state, which originates from the $N = 0$ Landau level. The factor of 4 is attributed to the spin and valley degeneracies, and the other factor of 2 comes from the two layers.(*38*) We have observed degeneracy lifting at a higher field of 100 mT (Supplementary Fig. 5). The data for our other sample, D2, where the onset of the QHE was observed at 5 mT is shown in Supplementary Fig. 6 and Supplementary Note 3.

As shown in Fig. 2(d), the Hall resistance reaches the first QHE plateau at $B = 2$ mT. This is lower than values reported for other graphene systems,(39) including suspended devices.(8) The quantum mobility, $\mu_q$, is expected to be even higher as it is defined by the onset of SdH oscillations corresponds to the value $10^7$ cm$^2$V$^{-1}$s$^{-1}$ according to the formula $\mu_q B = 1$. In Fig. 2(b), we can track the corresponding minimum in longitudinal resistance, $R_{xx}$, below 1 mT, as indicated by the dashed red line (see also Supplementary Fig. 10). The simultaneously measured $R_{xy}$ is shown at Fig. 3(e) with multiple plateaus developed above $B = 5$ mT. Such high mobility is correlated with a small residual concentration near the Dirac point (Supplementary Fig. 7 and Supplementary Note 4). Thus, the system is highly uniform, and it is expected to exhibit the FQHE at a relatively small magnetic field as the disorder caused by remote impurities is reduced, as will be discussed below.

It is possible that our estimation of $\mu_q$ is a lower bound, and other factors limit the observation of QHE plateaus below 2 mT in Device D5. The simplest explanation is that the development of well-defined plateaus is restricted to the regime $\mu_q B = 1$. Given our extracted quantum mobility $10^7$ cm$^2$V$^{-1}$s$^{-1}$, the condition for the onset of the Quantum Hall Effect is met precisely around $B = 2$ mT. Beyond this, the physical dimensions of the contacts may also contribute to the absence of plateaus at lower fields. The magnetic length is defined as $l_B = \sqrt{\hbar/eB}$, where $\hbar$ is the reduced Planck constant and $e$ is the elementary charge. At $B = 2$ mT, $l_B \approx 570$ nm, which is approximately half the width of our contacts. This suggests that as the magnetic field decreases and $l_B$ increases further, scattering between counter-propagating edge states within the metallic contact regions may prevent the formation of wider QHE plateaus.

By comparing devices with different channel widths, $W$, specifically, D1 and D3 with $W \leq 2$ µm, and D2, D4, and D5 with $W \geq 4$ µm, it is clear that an increase in channel width significantly enhances mobility. The field-effect mobility in device D5 can be evaluated via a transverse magnetic focusing experiment measured at $T = 15$ K as shown in Fig. 2(f). Here, longitudinal resistance as a function of perpendicular magnetic field and total concentration is shown with a clear QHE in the central regions, and several pronounced minima developed below 20 mT, the strongest of which is indicated by the green dashed curve. This minimum corresponds to a cyclotron radius $r_c$ of 1.75 *m*m which is half the distance between the nearest contacts. The transport mobility in a single graphene layer is given by $\mu_{tr} = \pi e r_c / \hbar\sqrt{\pi n_{1L}}$, where $n_{tot} = 2n_{1L}$. The transport mobility is approximately ≈ *4.7×10⁶* cm²V⁻¹s⁻¹ at $n_{tot} = 10^{10}$cm⁻². At room temperature, the field-effect mobility of all the DLG devices is close to *2× 10⁵* cm²V⁻¹s⁻¹ (Supplementary Fig. 1).(31) This value is primarily limited by extrinsic phonon scattering from the surface phonon polaritons of the hBN, rather than the intrinsic acoustic phonon scattering limit of graphene.(42) The intrinsic acoustic phonon resistivity, which is independent of carrier density, is approximately *30* Ω in graphene; however, the mobility we extract reflects the linear-in-$n_{1L}$ contribution to conductivity influenced by the substrate. Comparing D2 and D4, which both have wider channels and similar hBN spacer thicknesses, we noted that they have different types of ohmic contacts. The graphite ohmic contact only moderately enhances mobility but helps to avoid the formation of p-n junctions, which is beneficial for measurements under high magnetic fields. This suggests that disorder at the etched edges remains significant, and the short length of the graphite probe contacts provides limited suppression of edge disorder. However, the introduction of any disorder is detrimental to the enhancement of mobility. Our results indicate that the width of the graphene channel has a significant influence on its mobility, which is in agreement with previously reported data,(39), suggesting that increasing the channel width may further enhance overall device performance.

Due to mutual screening in the DLG structure, scattering is reduced, thereby enhancing mobility. Thinner hBN spacers provide enhanced mutual screening, leading to higher mobility. For devices with ultra-high mobility and ballistic transport in bulk, edge scattering emerges as the dominant scattering process, highlighting the critical role of channel width.

**Fractional Quantum Hall Effect**

Figure 3(a) shows the longitudinal resistance as a function of the Displacement field and the total concentration measured at the magnetic field of 2 T. The hole-doped region shows QHE at $v_{tot}$ =− *1, − 2, − 3, − 4*, where the spin, valley, and layer degeneracies are fully lifted. There are also several fractional quantum Hall effect plateaus on $R_{xy}$ with corresponding blue dashed lines, with the best quality observed at the total filling factor equal to − *10/3*, which can be understood in the composite fermion picture with asymmetric doping in both layers. By assuming two

Landau levels filled in one layer and the second LL is partially filled in the second layer, we have for $v_{tot}^{-10/3} = -\frac{10}{3} = -2 - (1 + 1/3)$. The observation of FQH states at such a low magnetic field is surprising in our system, as e-e interactions are crucial in forming FQH states, and the screening by the other graphene layer is expected to be unfavorable for forming fractionally charged quasiparticles. However, this is supported by the absence of FQH states at equal charge concentrations in both layers. Thus, a finite displacement field and a very low density of states are required to observe the FQHE.

Placing a graphite gate in proximity to graphene allows us to obtain an ultra-low disorder system, but at the same time, the Coulomb interaction reduces the FQH gaps by a factor of 4 or 5.(39) To show the advantage of our heterostructure for observing FQHE we studied the effect of screening on the gap size at low magnetic field at 3 T (Fig. 3c). The temperature dependences of longitudinal and transverse resistances were measured between $v_{tot} = [-4, -3]$. We observed plateaus at the filling factors of $v_{tot}^{-18/5}$ and $v_{tot}^{-10/3}$. They also show a robust zero-resistance state, enabling the measurement of the temperature dependence of the longitudinal resistance (see Supplementary Fig. 11). The activation dependence is then analyzed (inset in Fig. 3c) using a standard Arrhenius formula $lnR_{xx} = lnR_0 - \Delta/2k_BT$, where $R_0$ is a scaling prefactor, $k_B$ is the Boltzmann constant, and $\Delta$ is the activation gap to be determined from the best fit. We found that the state for $v_{tot}^{-10/3}$ corresponds to an activation gap of $0.18 \pm 0.01$ meV and $v_{tot}^{-18/5}$ demonstrates a smaller gap of $0.10 \pm 0.02\, meV$. In recent measurements (39) of the gaps in ultra-high mobility graphene, the filling factors of *4/3* and *8/5* were not observed; however, the largest gap for the filling factor *2/3* measured at 12 T was equal to 0.3 meV. Assuming a $\Delta \propto \sqrt{B}$ of the gap, our gap at 3 T extrapolates to 0.36 meV at 12 T. This suggests our FQH gaps are less affected by screening in the system than those in (39).

In conclusion, intercalating a few layers of hBN between two graphene layers facilitates the mutual screening of charged disorder from the substrates, thereby achieving ultra-high sample quality. In DLG, the quantum mobility exceeds $10^7$ cm$^2$V$^{-1}$s$^{-1}$. This allows observation of the onset of the QHE at 2 mT, and the FQHE at $B = 2\,T$. The determined gap size of the FQH states at $v_{tot}^{-10/3}$ at 3 T is equal to $0.18 \pm 0.01$ meV. This value demonstrates lower sensitivity to screening than devices that use proximity screening.(39) Our results show that the thinner the hBN spacer, the more significant the mobility enhancement. Simultaneously, due to reduced bulk disorder, boundary scattering becomes dominant, making mobility directly proportional to the channel width.

## Methods

### Sample Fabrication

The fabrication process begins with the preparation of the constituent layers: two graphene films, a thin hBN flake, two hBN flakes for encapsulation, and graphite flakes for the contacts. Each material is exfoliated on separate $SiO_2$/Si substrates to identify suitable flakes under an optical microscope. The hBN flakes, chosen for their thickness and uniformity, with no residuals, serve as both the substrate and the encapsulating layer for the graphene flakes. The thickness of the bottom hBN layer is carefully selected to be between 20 and 55 nm, balancing substrate flatness and device isolation.

The assembly of the layered structure is achieved through a dry transfer method using a poly(bisphenol A carbonate) (PC) film attached to a polydimethylsiloxane (PDMS) stamp. Thanks to its ability to pick up (120°C) and release (180°C) flakes at controlled temperatures, the PC film is pivotal for the sequential stacking of the hBN, graphene, and graphite layers. Initially, the PC/PDMS stamp picks up the top graphite flake. Subsequently, the stamp is aligned and brought into contact with the top dielectric layer, hBN. Then, a graphite flake is picked up using the same technique. Using graphite ensures that the contact resistance is minimized without introducing additional chemical doping or contamination, which is often a challenge with conventional metal contacts. Then, the first graphene layer is followed by the thin hBN separation layer, the second graphene layer, and finally, the bottom hBN flake. None of the graphene and hBN flakes are aligned crystallographically to avoid the formation of the side Dirac points. Throughout this process, care is taken to ensure clean interfaces between each layer, minimizing wrinkles and bubbles that could detrimentally affect device performance.

Following the assembly of the layered structure, electron beam lithography (e-beam lithography) is employed to pattern the device geometry and define the areas for the graphite contacts. Metal thermal evaporation was used to form Ti/Au contacts on the graphite edge.

### Transport Measurements

The devices D1, D2, D3, D4, and D5 were measured in an Oxford Instruments cryogen-free superconducting magnet system (Triton 500, 12 mK, 14 T), with the magnetic field applied perpendicular to the film plane. NF LI5650 lock-in is used to apply an AC with a 100 MOhm ballast resistor at a frequency of 17.777 Hz, and Keysight B2912A was used to apply voltages to the gates.

The device DG2 was measured in an Oxford Instruments cryogen-free superconducting magnet system (Heliox, 1.6 K, 14 T), with the magnetic field applied perpendicular to the film plane. NF LI5650 lock-in is used to apply an AC with a 100 MOhm ballast resistor at a frequency of 17.777 Hz, and Keysight B2912A was used to apply voltages to the gates.

**Data Availability:** Relevant data supporting the key findings of this study are available within the article and the Supplementary Information file. All raw data generated during the current study are available from the corresponding authors upon request.

Acknowledgments: The authors would like to thank the International Joint Lab of 2D Materials at Nanjing University for the support, and Yuanchen Co, Ltd (http://www.monosciences.com) for the Ultra-High-Mobility 2D Transfer System, which is one of the best transfer system to make van der waals heterostructures in the world. Geliang Yu acknowledges financial support from the National Key R&D Program of China (Nos. 2024YFB3715405, Nos. 2022YFA120470), the National Natural Science Foundation of China (Nos. 12004173, 11974169), and the Fundamental Research Funds for the Central Universities (Nos. 020414380087, 020414913201). K.S.N. is grateful to the Ministry of Education, Singapore (Research Centre of Excellence award to the Institute for Functional Intelligent Materials, I-FIM, project No.EDUNC-33-18-279-V12) and to the Royal Society (UK, grant number RSRPR 190000) for support. L.W. acknowledges the National Key Projects for Research and Development of China (Grant No. 2022YFA1200141 and 2021YFA1400400), National Natural Science Foundation of China (Grant No. 12074173), Natural Science Foundation of Jiangsu Province (Grant No. BK20220066 and BK20233001), Program for Innovative Talents and Entrepreneur in Jiangsu (Grant No. JSSCTD202101) and Nanjing University International Research Seed Fund.

**Author contributions:** G.Y. conceived the experiment. P.W., R.D., D.Z., Q.G., and

**Conflict of Interest:** Authors declare no competing interest.

**Supplementary Materials**

Supplementary Notes 1-4

Supplementary Figs. 1-11

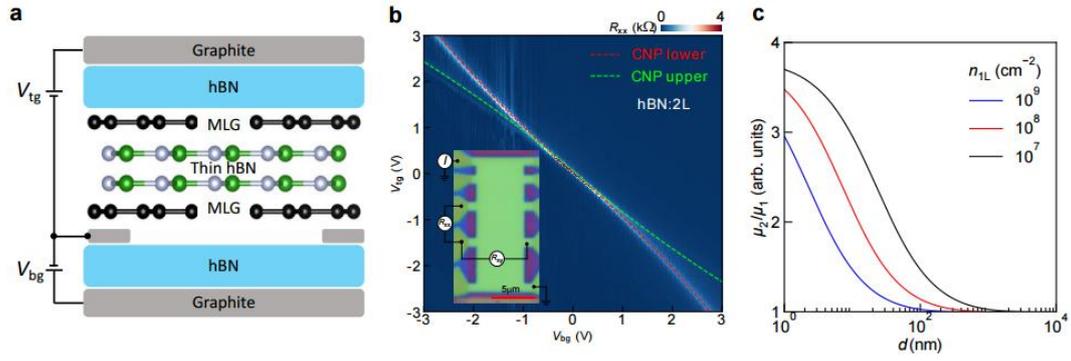

Figure 1: Sample's design and basic characteristics. (a) Schematic of the samples showcasing two layers of monolayer graphene (MLG) separated by a thin hexagonal Boron nitride (hBN) spacer and encapsulated between thicker hBN layers with dual graphite gates. The top gate voltage ($V_{tg}$) and the bottom gate voltage are applied between the graphite contacts to the graphene and the top and bottom gates, respectively. (b) For device D1, the resistivity $R_{xx}$ in the top gate voltage−bottom gate voltage ($V_{tg}$−$V_{bg}$) space in regions with 2L spacers. The red and green dashed curves represent the position of the charge neutrality point in the lower and upper graphene layers, respectively. The inset shows an optical image of the device with electrical connections, where $I$ is the current, $R_{xx}$ and $R_{xy}$ are the longitudinal and transverse resistances, respectively. The red scale bar is 5 μm. (c) The calculated mobility ratio ($\mu_2/\mu_1$) between the double- ($\mu_2$) and single-layer ($\mu_1$) structures as a function of the layer separation $d$, with an impurity concentration in a single layer ($n_{1L}$) at $10^7$ cm$^{-2}$ (black), $10^8$ cm$^{-2}$ (red), and $10^9$ cm$^{-2}$ (blue solid curve).

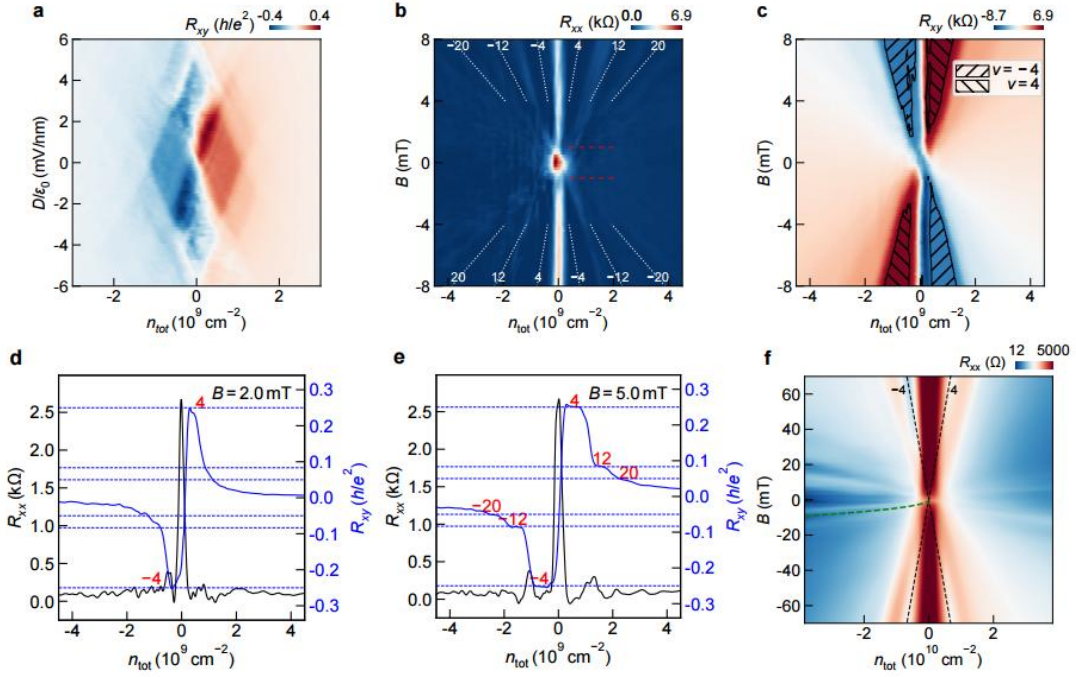

Figure 2: Quantum properties at small magnetic fields in device D5. (a) Transverse resistance $R_{xy}$ as a function of displacement field $D/\varepsilon_0$ and total concentration $n_{tot}$ at $B = 5$ mT and $T = 20$ mK, where $\varepsilon_0$ is the vacuum permittivity. (b) Longitudinal resistance $R_{xx}$ as a function of total concentration and magnetic field. The onset of the Shubnikov-de Haas oscillation is shown by the red dashed lines at $B = \pm 1$ mT. The white dashed lines show ($n_{tot} = \nu_{tot} eB/h$) the position of the $\nu_{tot} = \pm 4, \pm 12, \pm 20$ filling factors. (c) Transverse resistance $R_{xy}$ as a function of total concentration and magnetic field. The quantum Hall effect (QHE) measured at $D = 0$ V/nm. The dashed regions correspond to the $\rho_q \approx 7.3$ kΩ with an error of 2%. (d-e) The longitudinal and transverse resistances cuts from panels (b) and (c) at $D = 0$ V/nm for $B = 2$ mT and 5 mT. The horizontal dashed blue lines correspond to the plateaus for the total filling factors of $\nu_{tot} = \pm 4, \pm 12, \pm 20$. A developed plateau with corresponding zero longitudinal resistance is shown in (e) at $\nu_{tot} = \pm 4$. (f) Longitudinal resistance as a function of magnetic field and total concentration measured at $T = 15$ K. The black dashed lines correspond to the filling factors of $\nu_{tot} = \pm 4$. The green dashed line is the transverse magnetic focusing contribution from ballistic electrons in an individual graphene

layer and cyclotron radius of 1.75 mm.

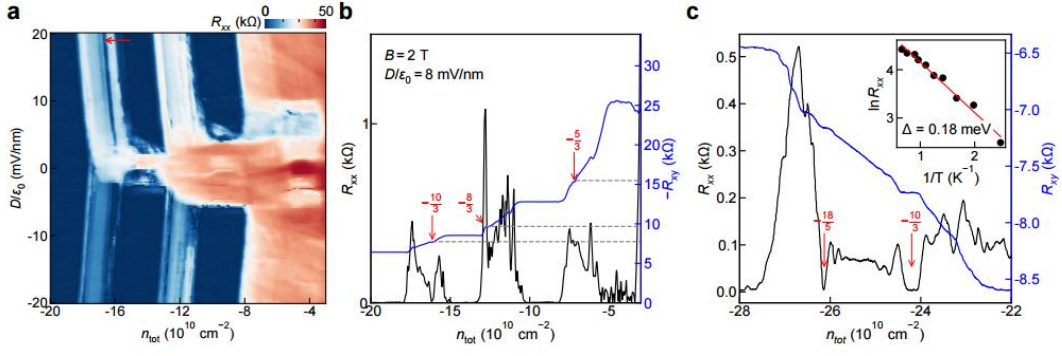

Figure 3: Fractional Quantum Hall (FQH) effect at low magnetic fields in device D5. (a) Longitudinal resistance $R_{xx}$ as a function of $D/\varepsilon_0$ and $n_{tot}$ at $B = 2$ T and $T = 15$ mK. The red arrow shows the zero-resistance state for the total filling factor of $\nu_{tot} = -10/3$. (b) Longitudinal and Hall resistances ($R_{xx}$ and $R_{xy}$) as a function of total concentration at 2 T. The gray dashed lines show the position of the FQH plateaus at $\nu_{tot} = -5/3, -8/3, -10/3$ filling factors, however, only the $\nu_{tot} = -10/3$ state has a corresponding zero longitudinal resistance. (c) Longitudinal and Hall resistances ($R_{xx}$ and $R_{xy}$) as a function of total concentration at 3 T. The red arrows show the position of the FQH zero-resistance states $\nu_{tot} = -10/3, -18/5$, where the gap size was measured. The inset shows the temperature dependence of the logarithm of the resistance. The red line is the best fit using equation $lnR_{xx} = lnR_0 - \Delta/2k_BT$, where $R_0$ is the reference resistance, $k_B$ is the Boltzmann constant, $T$ is the temperature.